\documentclass[11pt,twoside]{article}
\usepackage{./asp2014}

\aspSuppressVolSlug
\resetcounters

\begin{document}

\title{Developing Virtual Reality Activities for the Astro 101 Class and Lab}
\author{Gur Windmiller, Philip Blanco, and William F. Welsh}
\affil{San Diego State University, San Diego, California, USA; \email{gwindmiller@sdsu.edu}}

\paperauthor{Sample~Author1}{Author1Email@email.edu}{ORCID_Or_Blank}{Author1 Institution}{Author1 Department}{City}{State/Province}{Postal Code}{Country}
\paperauthor{Sample~Author2}{Author2Email@email.edu}{ORCID_Or_Blank}{Author2 Institution}{Author2 Department}{City}{State/Province}{Postal Code}{Country}
\paperauthor{Sample~Author3}{Author3Email@email.edu}{ORCID_Or_Blank}{Author3 Institution}{Author3 Department}{City}{State/Province}{Postal Code}{Country}

\begin{abstract}
We report on our ongoing efforts to develop, implement, and test VR activities for the introductory astronomy course and laboratory. Specifically, we developed immersive activities for two challenging "3D" concepts: Moon phases, and stellar parallax. For Moon phases, we built a simulation on the \textit{Universe Sandbox} platform and developed a set of activities that included flying to different locations/viewpoints and moving the Moon by hand. This allowed the students to create and experience the phases and the eclipses from different vantage points, including seeing the phases of the Earth from the Moon. We tested the efficacy of these activities on a large cohort (N=116) of general education astronomy students, drawing on our experience with a previous VR Moon phase exercise (\citet{Blanco2019}). We were able to determine that VRbased techniques perform comparably well against other teaching methods. We also worked with the studentrun VR Club at San Diego State University, using the Unity software engine to create a simulated space environment, where students could kinesthetically explore stellar parallax - both by moving themselves and by measuring parallactic motion while traveling in an orbit. The students then derived a quantitative distance estimate using the parallax angle they measured while in the virtual environment. Future plans include an immersive VR activity to demonstrate the Hubble expansion and measure the age of the Universe. These serve as examples of how one develops VR activities from the ground up, with associated pitfalls and tradeoffs.
\end{abstract}

\section{Virtual Reality for the Introductory Astronomy Laboratory}


Our small group at San Diego State University has been exploring the possible use of Virtual Reality (VR) to provide authentic, student-centered laboratory activities to teach difficult-to-grasp 3-dimensional concepts in general education astronomy lecture and lab classes. The hope is that by putting students in the driver’s seat and allowing them to control their vantage point, they will achieve a deeper understanding of these concepts. Our initial trials comparing VR with other teaching methods for Moon phases and eclipses show promise (\citealt{Blanco2019}; \citealt{Welsh2020}). This is in accord with the results of a similar Moon Phases study by \citet{Madden2020}. Here we report on a continuation of this work, and on the creation of a new VR laboratory activity on stellar parallax.

\section{Moon Phases and Eclipses}

A number of excellent 2D Moon phase simulators are available online, such as the app provided by the University of Nebraska, Lincoln. We wanted to create a much more immersive experience, with six degrees of freedom (“6-dof”), meaning 3 axes of rotation plus 3 directions of translation. These allow the student to look in any direction, and also change their viewpoint by moving around in a defined space (e.g. walking between the Earth and Moon). We adapted the free VR package \textit{Universe Sandbox 2} for this activity. Initially we tried keeping the Sun-Earth-Moon system to an accurate relative scale, but this proved awkward for educational purposes – the angular sizes of the Sun and Moon are too small compared to the full field of view. We therefore created a “closer Moon” to facilitate one’s ability to see Moon’s phase and shadow on the Earth, and to create an eclipse by grabbing this Moon using the HTC Vive’s hand controllers. 

As described in \citet{Blanco2019}, we used a “buddy system” to deploy this immersive activity, where each of a pair of students took turns. While one wore the headset, the other read the exercise instructions, recorded results, and monitored the headset view on a large 2D display. The VR room was located in the campus library (“BuildIT MakerSpace” area) so the students had access to the equipment most of the time. Based on feedback from our initial testing (\citealt{Blanco2019}), we created a pre-lab introductory video describing the headset and hand controllers. 

We administered pre- and post-VR experience quizzes to test students' understanding, and did the same with a "control" group taught by traditional methods. Assessment methods and results are reported in \citet{Welsh2020};  Figure \ref{Fig1} shows an example. In summary, we found that students who participated in the VR activity did at least as well as students who were taught with traditional methods, as found by \citet{Madden2020}. 
\articlefigure{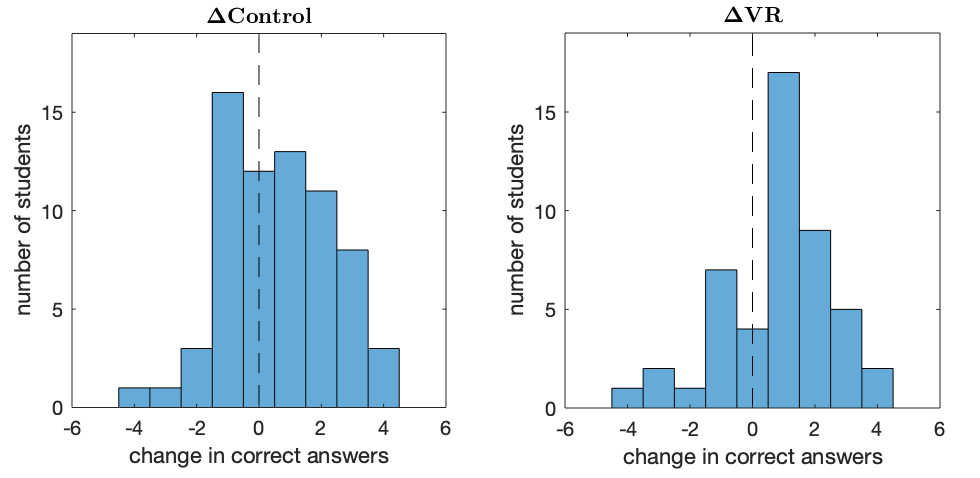}{Fig1}{The change in quiz score for students who did the Moon VR activity ($N=48$) versus a control group ($N=68$) taught with traditional methods. See \citet{Welsh2020}} 

\section{Stellar Parallax}

Having demonstrated that the Moon VR activity did not impede student learning, we chose stellar parallax as the next topic to investigate. This is a difficult activity to implement, first because of conceptual difficulties in students’ understanding of the motions and geometry, and second the instructor’s challenge of a creating an authentic 3D activity. In the past, we have set up objects at various distances in the laboratory, or outdoors, and provided instructions to students to observe the objects from two positions separated by a known baseline distance. However, the analogy to stellar parallax is difficult for students to conceptualize, and the apparent position change is confined to a straight line, which is unlike the motion of the Earth about the Sun. A VR simulation allows us to correct these deficiencies.

Since no off-the-shelf software existed to demonstrate stellar parallax in 3D, we teamed up with SDSU's student-run VR Club. We chose to simulate real asterisms (the Big Dipper and Cassiopeia), set against a backdrop of stars at infinite distance. Custom code was written by the students using the \textit{Unity} platform. We met weekly for a semester, providing guidance (and pizza) to the students. The learning went both ways – we explained the astronomy and math to the students, and the students explained optimal ways to achieve our goals, including better visualization ideas and thoughts on how to “gameify” the exercise.

The outcome of this collaboration was a two-part activity (Figure \ref{Fig2}). In the first part, students would play a game where the goal was to rank the distances to stars in an asterism, by moving from side-to-side to gauge the parallax. This provided an introduction to the second, quantitative activity, where the student was carried around the Sun on an Earth orbit, and their task was to trace out the apparent path of a nearby star relative to the background stars, using a hand controller pointer. Then, using an angle tool, they could measure the width of the resulting ellipse and use this to calculate a distance to the star. As with the Moon phases activity, scaling was necessary since real parallaxes of a fraction of an arcsecond are impossible to discern. Therefore, we artificially expanded Earth’s orbit such that observed parallaxes would be $5^{\circ}-10^{\circ}$ in the VR headset.

\articlefiguretwo{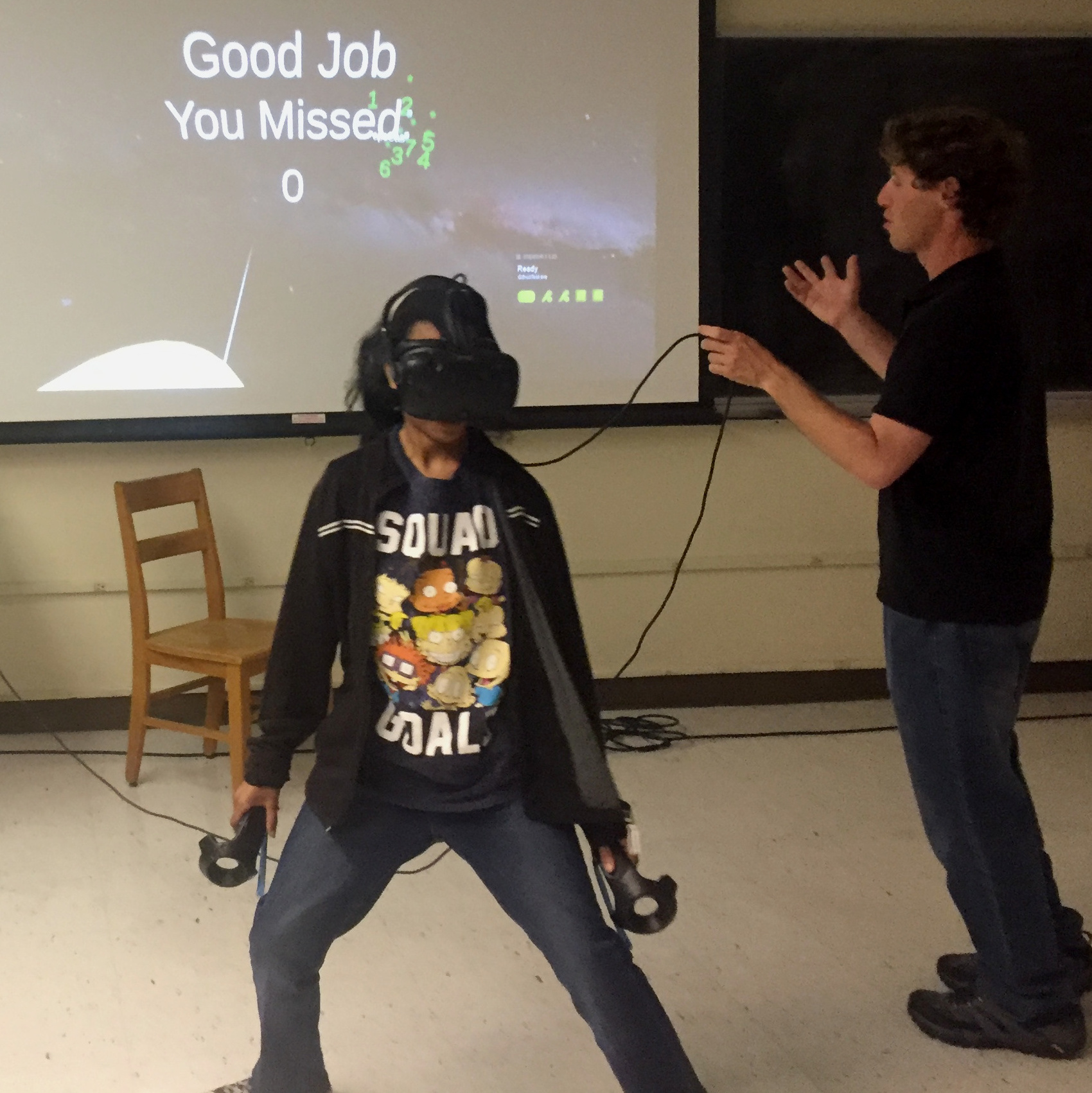}{Fig2right_850pix}{Fig2}{The stellar parallax activity was designed in two parts. Left: A game where students moved their viewpoint so that they could rank stars in an asterism in order of distance. Right: Students then used the hand controller to trace the apparent elliptical path of a star on the sky as they orbited the Sun along with the Earth.}

Most students were able to orient themselves and use the hand controllers easily. From an instructor’s standpoint, the biggest change from a traditional laboratory exercise where students follow instructions “in parallel”, was that they now had to wait their turn and perform the activity “in series”.  All students completed a worksheet where they calculated distances to the stars using the parallax angle they measured in the second part of the activity. (We did not count “missed guesses” from the first part against their grade.) In addition, we asked conceptual questions to test their depth of understanding. For example,  ``If we could make parallax measurements from Jupiter, would they be better or worse than what we can do from Earth?'' The vast majority of students performed well on these tasks. 

\section{Conclusions and Further Directions}

Producing VR activities is a lot of work! And although we wanted to keep astronomical scales authentic, this was not a good idea educationally. However, VR served very well in demonstrating the geometries of both Moon and stellar parallax activities. A key part of our methodology was to include time for unscripted activities to allow students to explore these difficult concepts on their own. While
learning outcome improvements are modest, part of that may be due to inadequate assessment tools. We do know that student enthusiasm for VR is very high.

Future activities include a demonstration of the Hubble expansion, where students can “fly” to different galaxies and construct their own Hubble diagram from any vantage point. In a draft version of this activity, we established reasonable sizes and number  of galaxies that allow students to collect useful quantitative datasets for offline analysis.  We invite interested readers to contact the authors for additional information, or to provide suggestions for additional activities. 









\acknowledgements The authors thank John Hood, Jr. for support, and the SDSU  Instructional Technology Services and Library for providing equipment and space for student testing. We also thank the members of the SDSU VR Club, especially Carter Andrews and Kain Shepard, for their indispensable assistance in writing code and providing valuable feedback on the stellar parallax activity



\end{document}